# Mutual modulation via charge transfer and unpaired electrons of catalytic site for superior intrinsic activity of N$_2$ reduction: from high-throughput computations assisted with machine learning perspective

Zheng Shu,[a] Hejin Yan,[a] Hongfei Chen,[a] Yongqing Cai[a*]

[a]Joint Key Laboratory of the Ministry of Education, Institute of Applied Physics and Materials Engineering, University of Macau, Taipa, Macau, China

## ABSTRACT

Electrocatalysts of nitrogen reduction reaction (NRR) have attracted ever-growing attention due to its application for renewable energy alternative to fossil fuels. However, activation of inert N-N bond requires multiple complex charge injection which complicates the design of the catalysts. Here via combining atomic-scale screening and machine learning (ML) methods we explore the rational design of NRR single-atom catalysts (SACs) supported by molybdenum disulfide (MoS$_2$). Our work reveals that the activity of NRR SACs is highly dependent on the number of unpaired d electrons of TM: "positive" samples with high activity favoring higher values while "negative" cases distributing at lower values, both varying with the doping conditions of the host. We find that the substitution of sulfur with boron can activate the intrinsic NRR activity of some TMs such as Ti and V which are otherwise inactive above pristine $MoS_2$. Importantly, among the various descriptors used in ML, the charged state of adsorbed TM plays a key role in donating electron to $\pi^*$ anti-bonding orbital of $N_2$ via the back-donation mechanism. Our work shows a feasible strategy for rational design of NRR SACs and retrieval of the decisive feature of active catalysts.

1. Introduction

As one of the most important raw materials in modern chemical industry[1,2], ammonia ($NH_3$) synthesized via Haber-Bosch approach under harsh condition[3] (200-250 bar, 400-500 °C) accounts for 1-2% of the total world's energy consumption[4]. Traditional bcc (110) surface Fe-based[5] and a close-packed hcp (0001) Ru-based[6] catalysts for nitrogen reduction reaction (NRR) are limited by high onset potential and low Faraday efficiency (FE) due to a coexisting high hydrogen evolution reaction (HER) activity[5-10]. The key difficulty of NRR is activation of inert N-N bond and converting $N_2$ to $NH_3$ ($N_2 + 6H^+ + 6e^- \rightarrow 2NH_3$) under mild conditions. Nowadays, electrochemical nitrogen reduction reaction (eNRR) attracts great interest because eNRR is environmental friendly and flexible with utilization of $H^+$ from water directly[11-15]. Single atom catalysts (SACs) are promising candidates for various catalytic tasks[16-19]. Compared with traditional trial and error method, computer simulation brings new perspectives to explore materials from atomic-scale sophisticated computations. In particular, density functional theory (DFT) calculations based on quantum mechanics play an important role in materials design and simulating physical chemical processes. However, computational burden still remains severely because DFT calculations need exhaustive electronic minimizations and ionic relaxations to identify potential configurational possibilities. Recently, data-driven method brought about great success in computer vision[20], natural language processing[21] and speech recognition[22]. With the inspiration and opportunities of machine learning (ML), amounts of related studies using ML-assisted method have emerged for prediction of formation energy of crystals[23], superhard compounds[24] and catalytic activity[25-31]. In general, material features should be properly selected for descriptor-to-property mappings in ML.

Ever since the discovery of graphene, two-dimensional (2D) layered materials have attracted much attention due to their unique mechanical, electronic and optical properties[32-35]. Amongst them, molybdenum disulfide ($MoS_2$) is a promising catalyst for HER and NRR, etc[16,36,37]. Recently, boron as a low-valent element has been exploited for NRR[38-41] because of its available empty orbitals for accommodating lone-pair electrons from $N_2$. Boron-doped graphene was shown with FE of 10.8% for NRR[42]. The boron can not only be active anchoring site but also substitutional doped element into the substrate[27]. Substitutional doping of boron can largely retain the host lattice unchanged and simultaneously induce tailored electronic structure which triggers the catalytic activity[28,29,43,44]. Kumar *et al.* demonstrated that boron-doped $MoS_2$ nanosheet is an effective strategy for tailoring the physical properties of $MoS_2$ and fabricated it successfully in experiment[43]. Li *et al.* showed di-boron pair doped $MoS_2$ monolayer exhibits superior catalytic activity for NRR by first-principles simulations[44]. Even so, the coupling between boron substitution and SACs with respect to their synergetic catalytic activity is still unknown and needs rational exploration.

Here, we reveal the decisive feature and the influence of the boron substitution on $MoS_2$ monolayer for NRR activity using first-principles calculations combined with ML-assisted method. By considering 26 different transition metals (TMs) anchored $MoS_2$ monolayer with doped boron cluster ($n$Bs-TM@$MoS_2$, n is the number of substitutional boron with sulfur, $n$ = 1-3), up to 104 potential catalysts are generated. Three key criterions, which can evaluate preliminary NRR performance, are used for labelling eligible catalysts to screen promising target. The analysis of feature importance inferred by ML is used to explore how the features influence the preliminary NRR activation, then thorough complete calculations for whole pathway of eligible catalysts were carried out. Amongst the various features, we have found the dominant descriptors, number of unpaired d electrons of TM and the charge transfer from TM, that govern the overall intrinsic NRR efficiency. From 104 potential catalysts, 1Bs-Os@$MoS_2$ possesses the best catalytic activity of NRR with the overpotential of 0.09 V and high selectivity up to 100% FE in theory.

## 2. Computational methodology

The configurational optimizations and energy calculations involved in this work were conducted with the Vienna ab initio simulation package (VASP)[45] using spin-polarized density functional theory (DFT). Generalized gradient approximation (GGA) with the Perdew--Burke--Ernzerhof (PBE) pseudopotential was adopted for the exchange correlation potential[46]. A monolayer $4 \times 4 \times 1$ supercell of $MoS_2$ was used as the substrate for catalysts with the kinetic cutoff energy of 450 eV. Usage of a larger supercell of $MoS_2$ ($5 \times 5 \times 1$ supercell) was proved to have minor effect within 20 meV[37]. Monkhorst-Pack scheme with a $3 \times 3 \times 1$ k-point mesh of the grid and a vacuum layer with thickness of 20 Å were adopted. All crystal geometric configurations were relaxed with the system energies and residual forces on each atom less than $1 \times 10^{-5}$ eV and 0.02 eV per Å, respectively. In addition, Grimme's semiempirical DFT-D3 method[47] was used to address the long-range van der Waals (vdW) force. The charge transfer of TM calculated by Bader charge method and the differential charge density calculations were implemented by VASPKIT[48] script. The thermodynamic stability of superior SACs was carried out by ab initio molecular dynamics (AIMD) simulations within the NVT ensemble at 400 K up to 10 ps with a time step of 2 fs. In our previous work, the solvent effect which was performed by implicit model VASPsol[49,50] was proved to play a minor role in NRR within 0.2 eV, which was also supported by other experimental and theoretical results[10,51-57]. Thus, the solvent effect was not taken into account in this work.

The change of Gibbs free energy ($\Delta G$) of elemental steps is the figure of merit for NRR catalytic activity. The $\Delta G$ in all NRR catalytic processes are calculated by computational hydrogen electrode (CHE) model developed by Nørskov[58]:

$$\Delta G = \Delta E + \Delta E_{ZPE} - T\Delta S \qquad (1)$$

where $\Delta E$ represents energy difference for products and reactants of each elementary step; $\Delta E_{ZPE}$ and $\Delta S$ calculated from vibrational frequencies are the changes of zero-point energy and entropy, respectively. $T$ is the temperature of system which is set to room temperature (298.15 K) in this work. The $\Delta E_{ZPE}$ and $T\Delta S$ can be obtained on the following equations:

$$E_{ZPE} = \frac{1}{2}\sum_i h\nu_i \qquad (2)$$

$$-TS = k_B T \sum_i \ln\left(1 - e^{-\frac{h\nu_i}{k_B T}}\right) - \sum_i h\nu_i \left(\frac{1}{e^{\frac{h\nu_i}{k_B T}} - 1}\right) \qquad (3)$$

where $h$, $k_B$, and $\nu_i$ represent Planck constant, Boltzmann constant, and vibrational frequencies of mode $i$, respectively. In the CHE framework, the potential-determining step (PDS) is the elementary step which has the most positive free energy change ($\Delta G_{max}$). Thus, the onset potential ($U_{onset}$) is determined by PDS calculated by: $U_{onset} = -\Delta G_{max}/e$ and the overpotential $\eta$ is determined by $\eta = U_{equillibrium} - U_{onset}$, where $U_{equillibrium}$ is -0.16 V in the NRR process. The standard values of $\Delta E_{ZPE}$ and $T\Delta S$ for gas $H_2$, $NH_3$ and $N_2$ are used in this work. For complex multielectron NRR reaction mechanisms, end-on pattern (only one end of $N_2$ is bonded to the active site) and side-on pattern (both two ends of $N_2$ are bonded to the active site) are both possible. Since the N₂ adsorption may be end-on or side-on, the corresponding Δ$G$(*N₂) is determined by the value whichever is smaller.

On the other hand, HER is the key competitive reaction for NRR, which can consume H⁺ and e⁻ to lead low FE of NRR. The activity of HER can be evaluated by:

$$\Delta G = \Delta E_{*H} + \Delta E_{ZPE} - T\Delta S_{*H} \qquad (4)$$

where $\Delta E_{*H}$ is the chemisorption energy of hydrogen and $\Delta E_{ZPE}$ is the change of zero-point energy between the hydrogen adsorbed state and the H₂. $\Delta S_{*H}$ can be approximated by $\Delta S_{*H} = -S_{H^2}/2$ [16,59], where $S_{H^2}$ is the entropy of gas hydrogen molecule. Thus, the $T\Delta S_{*H}$ is -0.21 eV under standard conditions[60]. For example, in the case of 2B$_S$-Cr@MoS₂, the vibrational frequency of *H has three eigenvalues 2150.41, 395.51 and 126.19 cm⁻¹, so the computed $E_{ZPE} = 0.17$ eV. To evaluate the selectivity of NRR compared with HER quantitatively, the theoretical Faradaic efficiency of NRR can be expressed according to Boltzmann distribution[18]:

$$f_{NRR} = \frac{1}{1+e^{-\frac{\delta G}{k_B T}}} \times 100\% \qquad (5)$$

where $\delta G$ is the Gibbs free energy difference between NRR and HER limiting-potential step, $k_B$ is Boltzmann constant and $T$ is room temperature. Furthermore, the details of ML models and data augmentation method could be found in ESI†.

## 3. Results and Discussion

## 3.1 Evaluation of Eligible Catalysts

Pristine MoS$_2$ and boron doped MoS$_2$ ($n$Bs-TM@MoS$_2$) monolayers are firstly calculated by DFT methods for training of ML and further screening. A full picture of computational flowchart is demonstrated in Fig. 1a. The atomic structures of boron doped MoS$_2$ with different boron replacement ratio considered in our work are shown in Fig. 1b. The lattice of doped MoS$_2$ is slightly distorted from pristine MoS$_2$ after doping different fractional boron atoms. In NRR, the activation of inert N≡N bond needs unoccupied and occupied orbitals of active site to accept electron density from N$_2$ and back donate electrons to π* orbital of N$_2$ for weakening N≡N bond simultaneously, which is termed as the π-backdonation mechanism[38].

Here three important criterions are used for screening eligible catalysts. Firstly, the change of free energy $\Delta G(*N_2)$ of N$_2$ chemisorption (* + N$_2$ → *N$_2$) is a key indicator for the appropriate interaction between gas N$_2$ and SAC[52-55]. A negative value of $\Delta G(*N_2)$, implying a spontaneous exothermic reaction which is favorable for the nitrogen reduction, is selected as the first criterion. The limiting potential which occurs in potential-determining step (PDS) is commonly used to evaluate the intrinsic activity of electrocatalysts[18,61]. It is defined as the highest free energy change at which the pathway is endothermic. In NRR reaction, the first hydrogenation process has been identified as the most common PDS for NRR catalysts[28,53,54]. Therefore, herein the $\Delta G$ of the first hydrogenation smaller than 0.55 eV is adopted as the second criterion for screening.

The DFT calculated $\Delta G$ of N$_2$ adsorption and the first hydrogenation of 104 candidates can be acquired in Tab. S1†, and the scatterplot of $\Delta G$(*N$_2$) versus $\Delta G$(*N$_2$ → *N$_2$H) for all the candidates are plotted in Fig. 1c and those cases appearing at the bottom left of the figure meet the abovementioned two criterions and thus used for further evaluation. It should be noted that $\Delta G$(*NH$_2$→*NH$_3$) was considered as an important process in some works[27,37]. However, this *NH$_2$ to *NH$_3$ reaction may not occur in the side-on pathway where replaceable process is *NH$_2$-*NH$_3$ to *NH$_3$. Many catalysts have shown superior activity by side-on pathway without the process of *NH$_2$ to *NH$_3$[39,54]. Hence, in our current work, the $\Delta G$(*NH$_2$→*NH$_3$) wasn't selected to be a rule for the judgement of eligible NRR catalyst.

The third screening rule is related to the selectivity of NRR. The free energy change of hydrogen adsorption $\Delta G$(*H) is examined because strong adsorption of atomic H would cover the surface of catalyst and suppress N$_2$ absorption. Hence the condition of $\Delta G$(*N$_2$) < $\Delta G$(*H) ensures a preferred N$_2$ absorption than hydrogen absorption. The $\Delta G$(*H) of residual candidates can be acquired in Tab. S2†. Fig. 1d gives a comparison of the energetics for adsorptions of atomic H and molecular N$_2$ for those pre-screened candidates. The samples which are located in upper-left region where $\Delta G$(*N$_2$) < $\Delta G$(*H) are eventually labeled as the eligible catalysts (i.e. energy favorable for N$_2$ adsorption, the first hydrogenation, and inhibition for hydrogen absorption) for binary classification and feature importance analysis using ML algorithms.

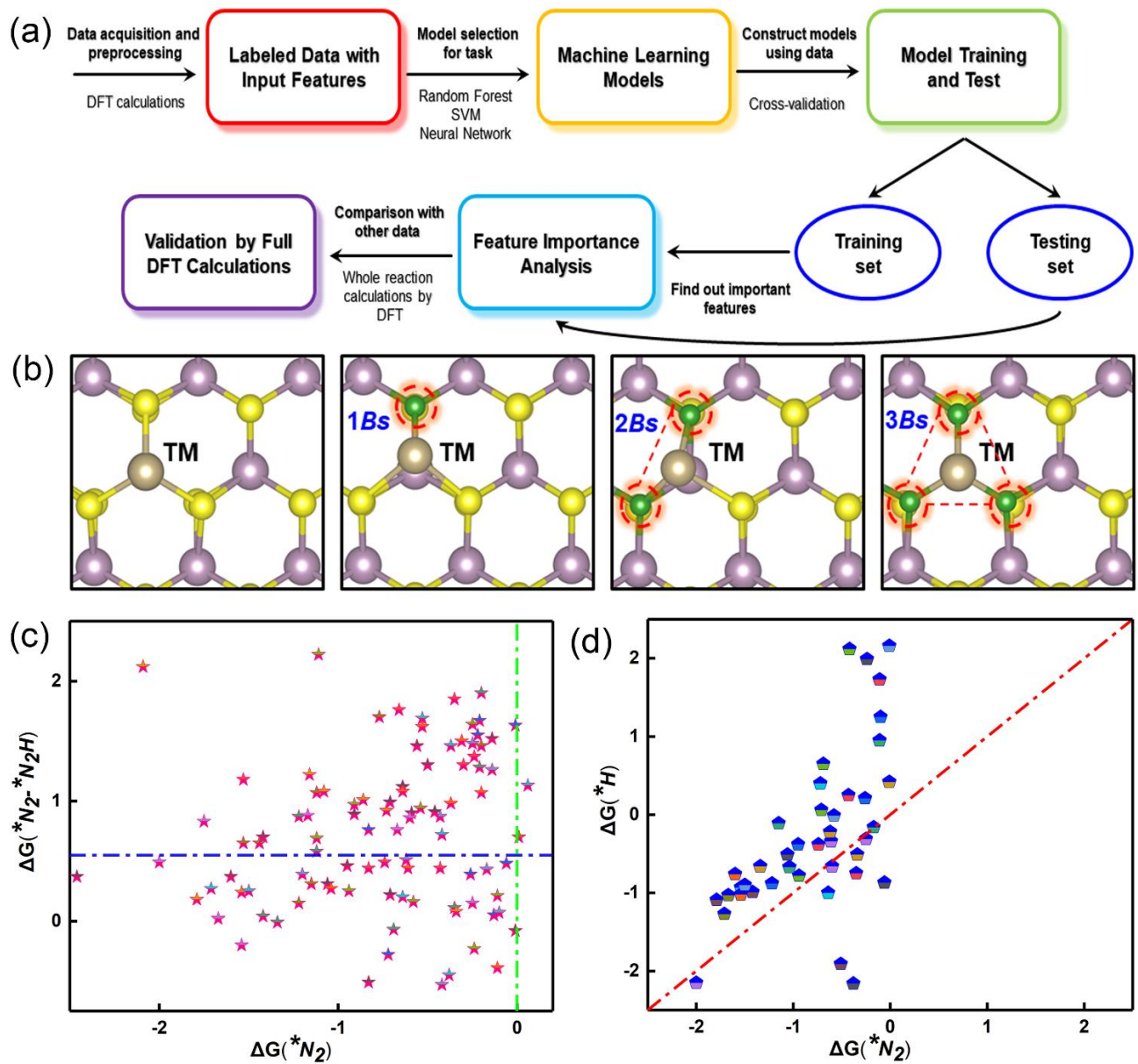

**Fig. 1** (a) The computational flowchart of DFT calculations combined with machine learning. (b) The local atomic structures of TM@MoS$_2$, 1Bs-TM@MoS$_2$, 2Bs-TM@MoS$_2$ and

3Bs-TM@MoS$_2$, respectively. The scatterplot of (c) change of free energy $\Delta G$ (*N$_2$) versus $\Delta G$ (*N$_2$ → *N$_2$H) and (d) $\Delta G$ (*N$_2$) versus $\Delta G$ (*H). The green, yellow, purple and grey balls represent B, S, Mo and transition metal (TM) atoms, respectively.

3.2 Feature Engineering and Importance Analysis.

The purpose of ML is through a binary classification task of eligible NRR SACs to preliminarily evaluate catalytic activity. Since the activity of catalyst is highly related to near-neighbor environment of active site[62], so our ML models are constructed on the basis of the following 11 descriptors: Mendeleev number ($Z$) and atomic radius ($R$) of TM, electronegativity ($\chi$) and ionization energy ($E_I$) of TM, the amounts of valence electrons ($N$) and unpaired electrons in d subshell ($N_{ie-d}$) of TM, formation heat of TM ($\Delta H_f$), charge transfer of TM ($\delta$), mean bond length between TM and sulfur atom ($L_{TM-S}$), mean bond length between TM and boron atom ($L_{TM-B}$), fraction of boron substitution ($f_B$).

Those features are summarized and used as chemical and structural representations of the active species and corresponding local coordination environment. In NRR, a positive charged TM active site is beneficial for suppressing H adsorption and boosting N$_2$ fixation to enhance the NRR performance. Therefore, the Bader charge ($\delta$) of the active site is involved. Moreover, the number of unpaired d electrons ($N_{ie-d}$) of TM is also included as a unique descriptor for NRR. The reason is that those half-filled d orbitals would not only can gain electrons from $N_2$ but also potentially donate electrons back to antibonding orbital of $N_2$. This "acceptance-donation" π-backdonation mechanism is beneficial for the activation of the N-N bond. The TMs considered in this work are compiled in Fig. 2a.

Next, we conduct various ML methods like decision tree (DT), random forest (RT), fully connected neural network (FCNN) by 4-fold cross validation method to cross check the generalization of different ML models. The schematic diagram and results on different ML models of 4-fold cross-validation are shown in Fig. S1† and Fig. 2b. In order to get the desirable result on specific dataset by FCNN, an appropriate design is required. The basic schematic architecture of FCNN is shown in Fig. 2c. FCNN is a hierarchical representation method via learning successive layers. Here, our designed FCNN was composed of an input layer with 11 neurons, two hidden layers with similar amounts of neurons as input layer followed by relu (x) activation function, and an output layer with one neuron followed by sigmoid (x) function for probabilistic output (0 ~ 1). Dropout layer was introduced for each hidden layer to ease overfitting. The Adam optimizer was used for gradient descent algorithm to update parameters of FCNN. The output layer of FCNN is designed for quantifying the probability of the candidate as an eligible catalytic activity represented by "1" or non-eligible activity denoted by "0". We found that the FCNN method possesses the best generalization whose mean accuracy is up to 92.31%, which is much higher than other algorithms (see Fig. 2b).

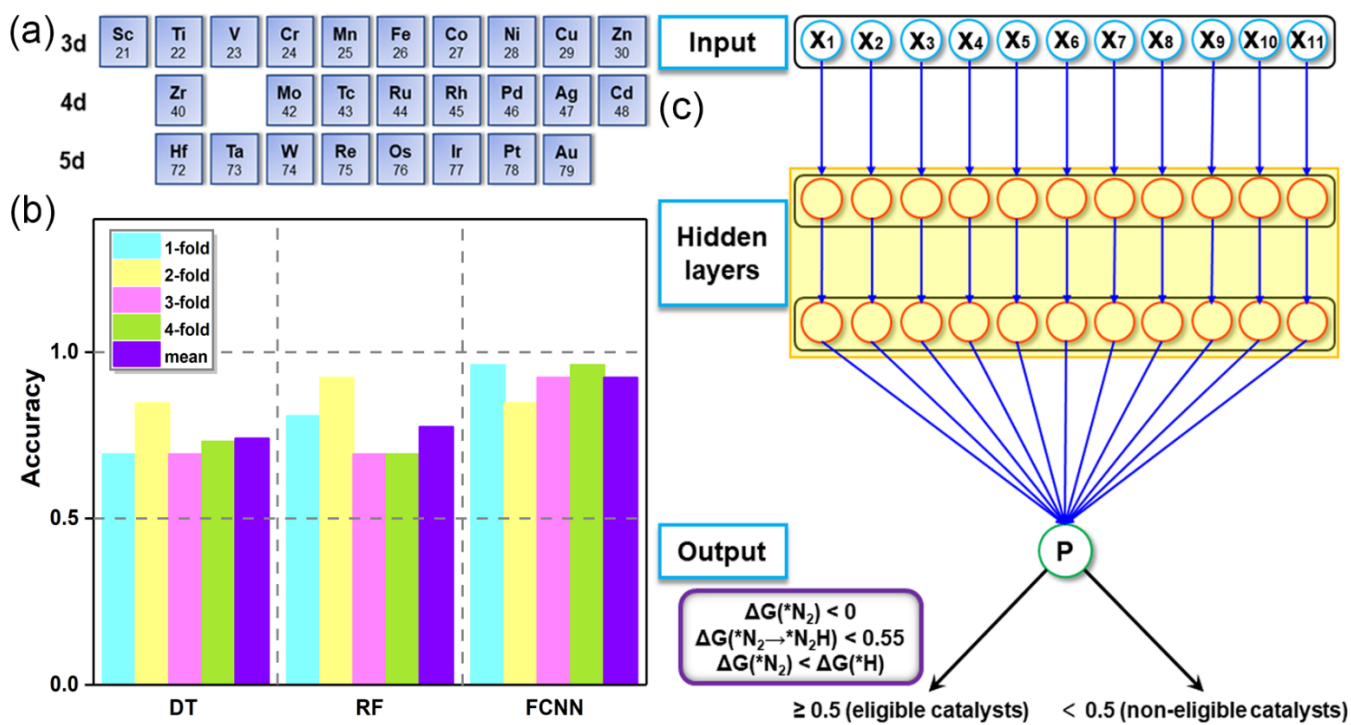

**Fig. 2** (a) Considered TM atoms in this work. (b) The mean accuracy of different ML models using 4-fold cross-validation. The fully connected neural network (FCNN) is the best model with the highest generalization. (c) The schematic architecture of FCNN with simplified neurons and 11 neurons are chosen as input features ($Z$, $\Delta H_f$, $R$, $\chi$, $E_I$, $N$, $N_{ie\text{-}d}$, $\delta$, $L_{TM\text{-}S}$, $L_{TM\text{-}B}$ and $f_B$). It should be noted that in the real model each neuron connects with all other neurons of neighboring layers to keep full-connected.

The reason of the other algorithms with a relatively lower accuracy compared to FCNN is further explored. The feature-to-feature heat map inferred by Pearson's correlation coefficient is presented in Fig. 3a, which shows the correlation between different features. The decision process of decision tree (DT) and the feature importance of random forest (RF) are visualized in Fig. S2 and S3†. The number of valence electrons $N$ is the first split feature which is thus the most important feature in DT, and it is the most important feature in RF. Interestingly, according to the DT and RF algorithms, the feature of the number of unpaired d electron $N_{ie-d}$ is less important. In particular, the DT can fit the training set up to 100% accuracy without the $N_{ie-d}$ descriptor. Similarly, according to RF deduction, the $N_{ie-d}$ is the second to last feature with the accuracy loss of 0. It implies that the $N_{ie-d}$ feature has no effect on prediction accuracy anymore, so $N_{ie-d}$ is an ignorable feature for eligible NRR catalysts. In contrast, the value of feature importance of $N_{ie-d}$ in FCNN model is 0.0623, as shown in Fig. 3b, ranked as the moderately important feature and much higher than that in DT and RF. This different hierarchy may be the reason that DT and RF have poor generalization than FCNN on test data. Therefore, we reveal that the number of half-filled d orbitals in TM plays an important role, and predictions excluding it may have uncertain deviations. In addition, the feature $f_B$ (the 11th feature in Fig. 3b) occupies dominant place in the FCNN model whose importance surpasses far from other features. It implies that boron substitutions in $MoS_2$ plays an important role for NRR activity.

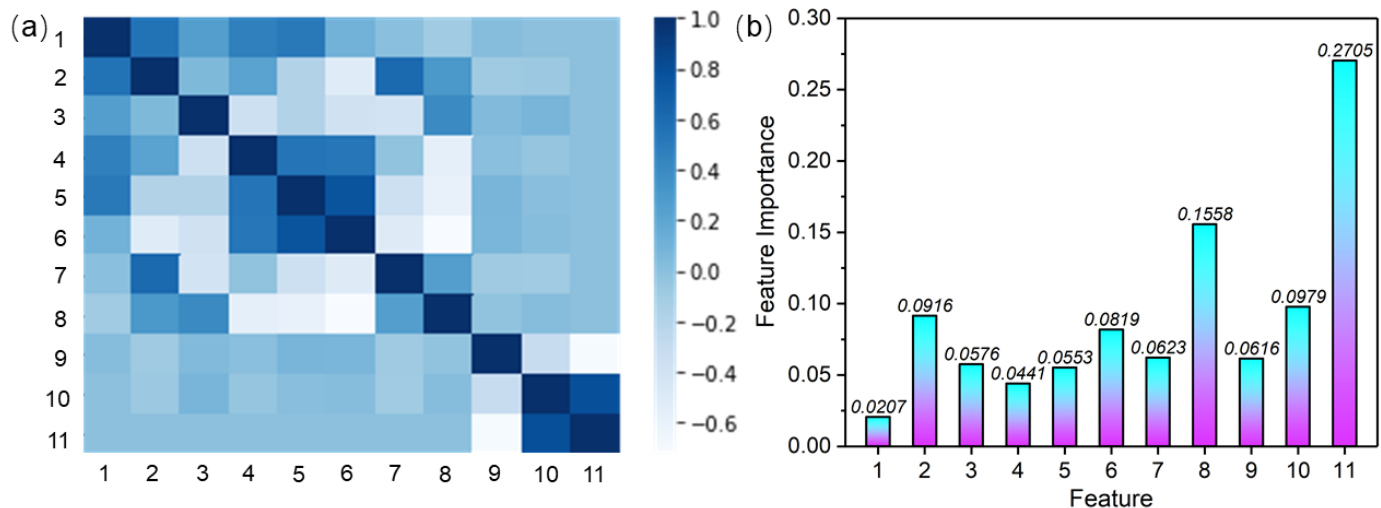

**Fig. 3** (a) Heat map inferred by Pearson's correlation coefficient. (b) The ranking of most important features of FCNN. Feature 1 to 11 represent $Z$, $\Delta H_f$, $R$, $\chi$, $E_I$, $N$, $N_{ie-d}$, $\delta$, $L_{TM-S}$, $L_{TM-B}$ and $f_B$, respectively.

### 3.3 Activation of N$_2$ and Whole Reaction

According to our simulation, the 3Bs-TM@MoS$_2$ cases were ruled out because the anchored TM atom and substituted B atoms strongly distort and deviate from original positions via molecular dynamics simulation. The remaining eligible catalysts are then passed for thorough complete NRR calculations. The calculated vibrational frequencies, $\Delta E_{ZPE}$ and $\Delta S$ of 2Bs-Mn@MoS$_2$, as the representative of the calculation, are listed in Tab. S3†. Herein we use the metrics of $\Delta G_{max} < 0.55$ eV as the criteria for screening promising catalysts. The detailed values of $\Delta G_{max}$ for all the eligible candidates are compiled in Tab. S4†.

We have conducted statistical analysis of all the promising candidates and obtained the probability density for the "positive" cases with satisfying $\Delta G_{max}$ and "negative" cases with unsatisfying large $\Delta G_{max}$. The results are presented as the violin plots of all the candidates (see Fig. 4a and b). As shown in Fig. 4a, for a specific structure disregarding the boron content, those favorable cases distribute at regions of higher $N_{ie-d}$ than the unfavorable TMs, indicating that a higher number of unpaired d electrons of a TM seems to have a higher probability of inducing promoted NRR activity. The introduction of boron atoms makes the distributions of eligible TMs more dispersive. For $n$=1 and 2 ($n$, the number of substitutional boron atoms), the peak positions and values of the probability curve corresponding to the favorable samples increase while those of unfavorable samples decrease with increasing $n$. The underlying reason is that the $B_S$ defect in $MoS_2$ induces p-type doping and holes (see DOS in Fig. S4†), thus a higher $n$ requires more d electrons ($N_{ie-d}$) from the TM to compensate the additional holes.

Interestingly, according to the statistics shown in the violin plot (see Fig. 4b), we find that a higher charged TM (~ +2) would be more likely to have a poorer performance than those lower charged TM (~ +1). This is reasonable since highly positive charged TMs are unfavorable for NRR activity as donation of electrons to $N_2$ is suppressed and thus being counterproductive for the "acceptance-donation" principle. Normally the NRR catalysts with $\Delta G_{max}$ < 0.4 eV (overpotential < 0.24 V) are considered as superior activity. Amongst the various candidates, the 1Bs-Ti@$MoS_2$, 1Bs-V@$MoS_2$, 1Bs-Os@$MoS_2$, 2Bs-

Mn@MoS$_2$ are screened by DFT calculations for potential superior NRR activity with $\Delta G_{max}$ of 0.37, 0.33, 0.25 and 0.39 eV, respectively. In particular, the 1Bs-Os@MoS$_2$ has the best performance with $\Delta G_{max}$ of 0.25 eV, showing a better performance than other studies[37,51,52]. The reaction processes with most energy favorable pathway are shown in Fig. 4c-f. The PDS of 1Bs-Ti@MoS$_2$ and 1Bs-V@MoS$_2$ via enzymatic pathway is the *NH-*NH$_2$ → *NH$_2$-*NH$_2$ process, and the PDS of 1Bs-Os@MoS$_2$ and 2Bs-Mn@MoS$_2$ via distal pathway is found to be the first hydrogenation process.

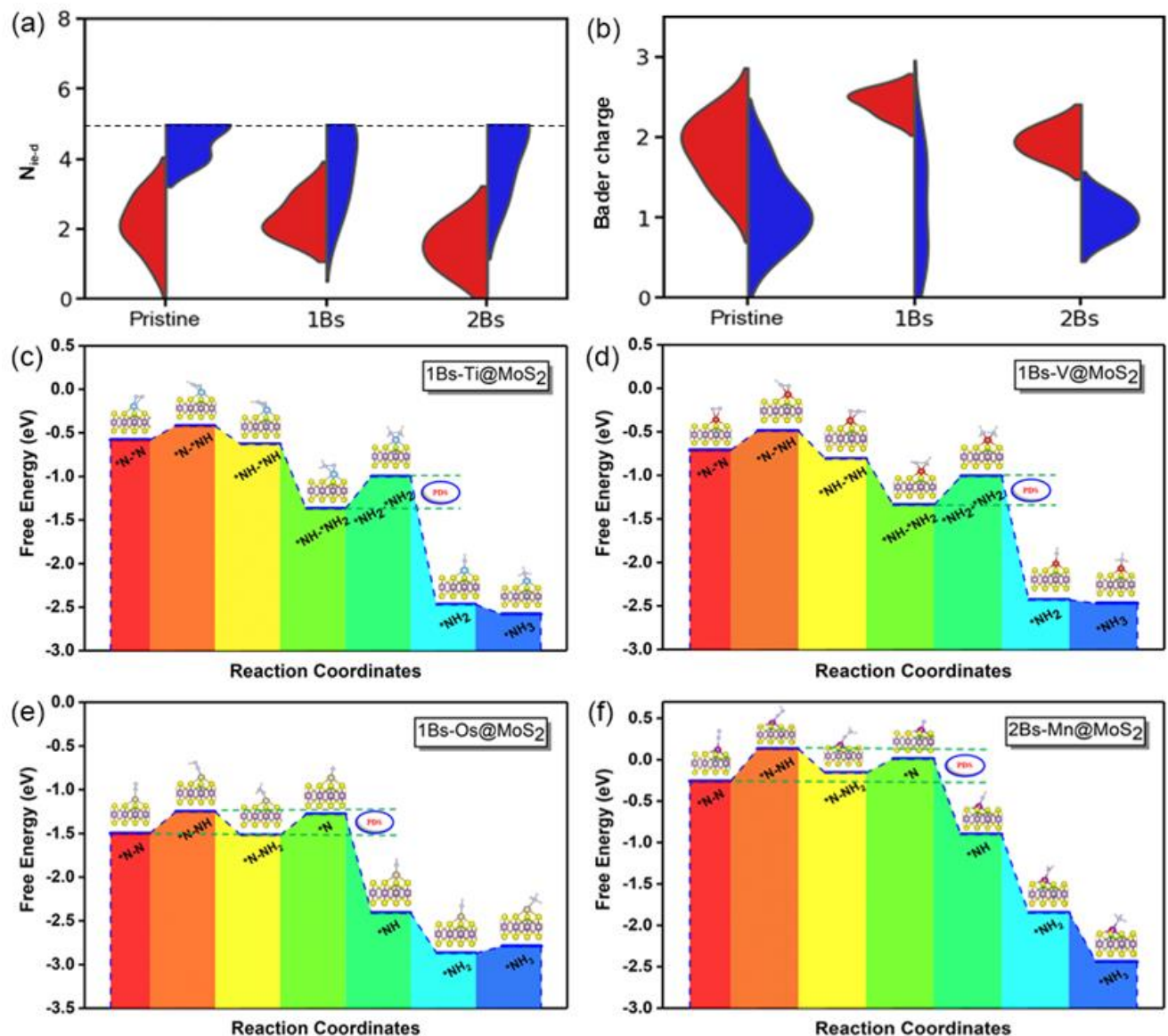

**Fig. 4** Violin plot for the influence of (a) the number of unpaired d electrons of TM atom, $N_{ie\text{-}d}$ and (b) Bader charge to NRR activity, respectively. The right blue part of violin plot represents the density distribution of promising catalysts while the left red part are unfavorable. Change of Gibbs free energy for the whole NRR process of (c) 1Bs-Ti@MoS$_2$, (d) 1Bs-V@MoS$_2$, (e) 1Bs-Os@MoS$_2$ and (f) 2Bs-Mn@MoS$_2$, respectively.

To get physical insight for NRR catalysts, the density of states (DOS) is analyzed. The DOS of TM active site and $N_2$ are shown in Fig. 5. The projected DOS as shown in Fig. 5a indicates that the d subshells are partially occupied and highly extended, especially true for Os. These d orbitals are resonant with the antibonding states of adsorbed $N_2$ (see Fig. 5b), facilitating the "acceptance-donation" π-backdonation mechanism. For instance, the PDOS of $N_2$ absorbed on 1Bs-Os@$MoS_2$ is highly overlapping with the PDOS of d orbital of Os atom which has dense states at Fermi level, thus favoring charge transfer with adsorbed $N_2$.

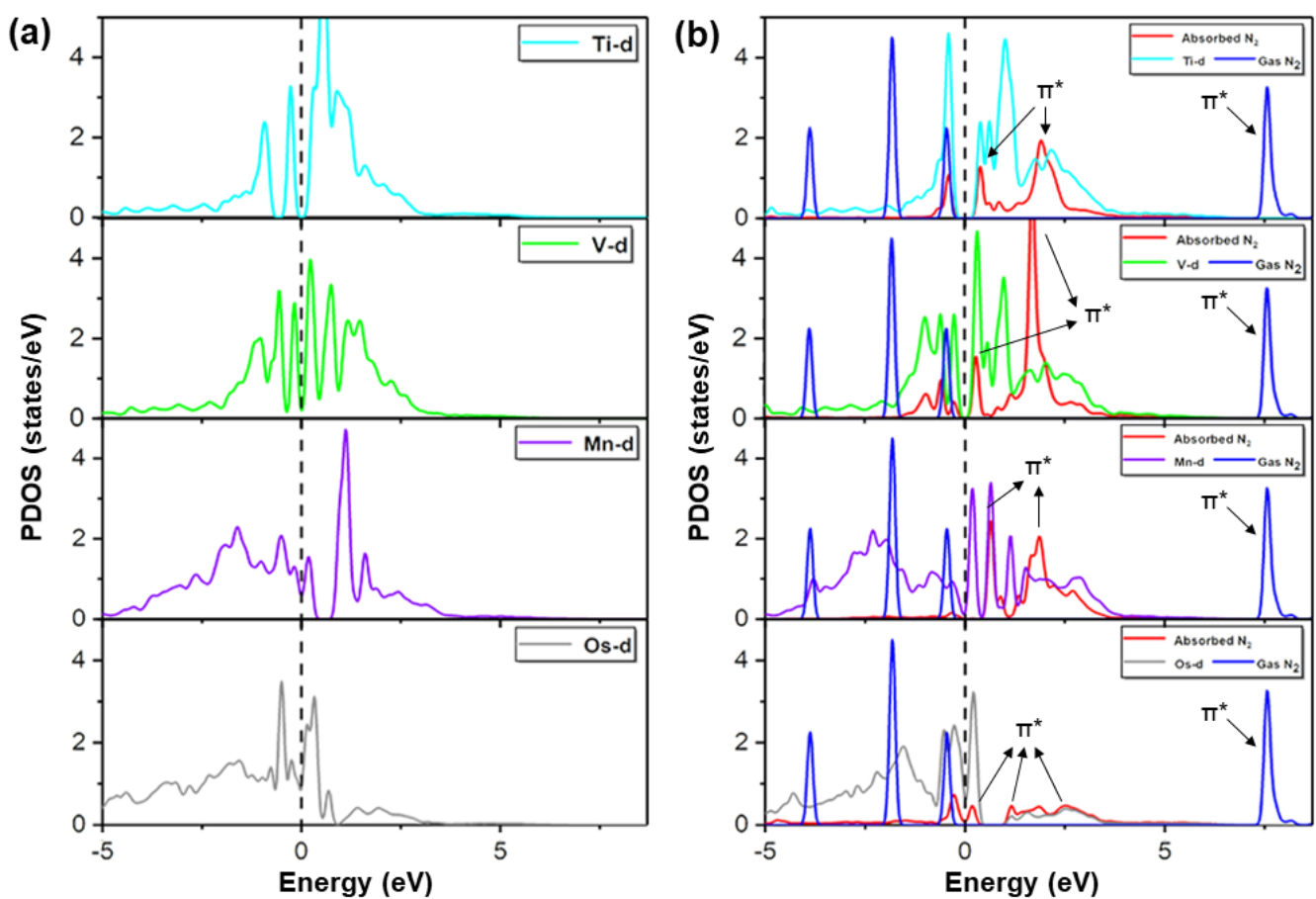

**Fig. 5** (a) Partial density of states (PDOS) of d orbital of metal atoms as active sites in the systems of 1Bs-Ti@MoS$_2$, 1Bs-V@MoS$_2$, 1Bs-Os@MoS$_2$ and 2Bs-Mn@MoS$_2$. (b) PDOS of isolated N$_2$ and absorbed N$_2$ compared with PDOS of d orbital of metal atoms after absorbing N$_2$. The Fermi level represented by dashed lines is set to zero.

The differential charge densities and ab-initio molecular dynamics (AIMD) simulations up to 10 ps with a time step 2 fs at 400 K of four superior catalysts are presented in Fig. S5-8†. It is noted that the active site and boron substitution still stay in original positions (Fig. S5-8b†). The energy and temperature (see Fig. S5-8c†) are oscillating around certain constant values, indicative of high stability of four catalysts. With increasing the fraction of replacing boron, the binding energy of TM increases accordingly. This implies that the boron substitution can promote the binding of TM. The comparison between different fraction of boron substitution for 1Bs-Ti@MoS$_2$, 1Bs-V@MoS$_2$, 1Bs-Os@MoS$_2$ and 2Bs-Mn@MoS$_2$ are shown in Fig. S9†. However, for $n=3$, the thermodynamic stability of configuration will become deteriorate. Therefore, a higher boron concentration is unfavorable due to the issue of thermodynamic stability though it may offer even stronger binding for TM species. The possible reason for this instability could be due to the increased mismatching of atomic radius between boron and sulfur. Further increasing the boron dopants would create somehow clusters-like boron structure embedding in MoS$_2$, and those clusters would tend to reconstruct and deviate from the planar structure.

**Table 1** The calculated Faradaic Efficiency (FE) values for all promising catalysts ($\Delta G_{max}$ < 0.55 eV).

|    | Mo@MoS$_2$ | Tc@MoS$_2$ | W@MoS$_2$ | Re@MoS$_2$ | Os@MoS$_2$ | 1Bs-Ti@MoS$_2$ |
|----|------------|------------|-----------|------------|------------|-----------------|
| FE | 99.98%     | 99.98%     | 100%      | 100%       | 100%       | 0               |
|    | 1Bs-V@MoS$_2$ | 1Bs-Mo@MoS$_2$ | 1Bs-Tc@MoS$_2$ | 1Bs-Ru@MoS$_2$ | 1Bs-W@MoS$_2$ | 1Bs-Re@MoS$_2$ |
| FE | 0          | 0          | 1.39%     | 0          | 100%       | 100%            |
|    | 1Bs-Os@MoS$_2$ | 2Bs-Cr@MoS$_2$ | 2Bs-Mn@MoS$_2$ | 2Bs-Tc@MoS$_2$ |  |  |
| FE | 100%       | 0.06%      | 0.09%     | 0          |            |                 |

Finally, the selectivity of NRR in comparison to HER was further calculated for all promising catalysts (see Table 1). It is revealed that 1Bs-Os@MoS$_2$ achieves a high theoretical FE of 100%. Although 1Bs-Ti@MoS$_2$, 1Bs-V@MoS$_2$ and 2Bs-Mn@MoS$_2$ show super intrinsic NRR activity, the more favorable HER activity may inhibit the NRR activity. However, 1Bs-Ti@MoS$_2$ and 1Bs-V@MoS$_2$ are superior catalysts for HER with $\Delta G(*H)$ close to zero (-0.01 eV and 0.06 eV, respectively). In addition, Mo@MoS$_2$, Tc@MoS$_2$, W@MoS$_2$, Re@MoS$_2$, Os@MoS$_2$, 1Bs-W@MoS$_2$ and 1Bs-Re@MoS$_2$ also have very high FE with desirable NRR activity of low overpotential, which are desired to be further evaluated experimentally.

# 4. Conclusion

In summary, the atomic scale DFT calculations combined with ML-assisted method are performed for rational design of NRR SACs. We reveal that the DFT and ML calculations provide mutual corroboration that the number of unpaired d electrons of TM plays the key role in nitrogen reduction process. In addition, the charge transfer from TM is also an important criterion for NRR, and appropriate positive charged TM can promote NRR activity and suppress hydrogen absorption. We found that the proper addition of substitutional boron atom on $MoS_2$ can activate intrinsic NRR activity of some metal sites compared with pristine $MoS_2$. After extensive screening, 1Bs-Os@$MoS_2$ is predicted to a superior NRR catalyst with overpotential of 0.09 eV and high theoretical FE of 100%. This can be attributed to the strong coupling between absorbed $N_2$ and the d orbitals of transition metal. Moreover, we uncover the key role of the host such as doping conditions in affecting the ideal features of appropriate SACs. In this work, a framework which can facilitate the discovery of more potential SACs is established for a rational design of SACs via first-principles calculations combined with ML approach.

## Author contributions

Zheng Shu: investigation, and writing original draft. Hejin Yan and Hongfei Chen: review & data analysis. Yongqing Cai: conceptualization, writing, review & editing, supervision, project administration, and funding acquisition.

## Conflicts of interest

There are no conflicts to declare.

## Acknowledgements

This work is supported by the University of Macau (SRG2019-00179-IAPME, MYRG2020-00075-IAPME) and the Science and Technology Development Fund from Macau SAR (FDCT-0163/2019/A3), the Natural Science Foundation of China (Grant 22022309) and Natural Science Foundation of Guangdong Province, China (2021A1515010024). This work was performed in part at the High Performance Computing Cluster (HPCC) which is supported by Information and Communication Technology Office (ICTO) of the University of Macau.